\documentclass{JAC2003}
\usepackage{graphicx}
\usepackage{booktabs}

\begin{document}
\title{INVARIANT CRITERION FOR THE DESIGN OF MULTIPLE BEAM PROFILE
EMITTANCE AND TWISS PARAMETERS\\ 
MEASUREMENT SECTIONS
\vspace{-0.6cm}}

\author{V.Balandin\thanks{vladimir.balandin@desy.de}, 
W.Decking, N.Golubeva \\
DESY, Hamburg, Germany}

\maketitle

\begin{abstract}
\vspace{-0.1cm}
In this and in the accompanying paper~\cite{PartII},
we introduce and give examples of applications of an optimality
criterion which can be used for the design and comparison
of multiple beam profile emittance and Twiss parameters 
measurement sections and which is independent from
the position of the reconstruction point.
\end{abstract}

\vspace{-0.2cm}
\section{INTRODUCTION}

\vspace{-0.1cm}
The standard approach to determine the transverse beam parameters 
at some location in a transport line is to measure, at first, the sufficient 
number of the beam sizes, then, using the known optics model between 
reconstruction point and measurement states, to find an approximation 
to the beam matrix (typically, by applying least squares fit), and, 
finally, to extract emittance and Twiss parameters from the 
approximation to the beam matrix obtained at the previous step.
The principal point of this procedure is the question of accuracy,
i.e. the question of the impact of the errors in the beam size measurements
on the precision of the reconstruction of the beam parameters.
Even though, in each particular situation, the errors of the reconstruction
of the emittance and Twiss parameters can be evaluated using a 
Monte-Carlo simulations, the numerical calculations alone
can not clarify all questions connected with the problem of
designing of a ``good measurement system".
For example, the question, if a $n$-cell measurement
system reaches an optimal performance when its
design Twiss parameters are cell periodic and the cell phase
advance is a multiple of $180^{\circ}$ divided by $n$,
is still a matter of controversy, though there is a considerable
amount of the numerical investigations of this problem 
made by different authors.
Thus an analytical criterion (even simplified),
which can provide a more or less general view on the problem
of errors in the beam parameter measurements and can also
guide more detailed numerical optimizations,
is still desirable.

Unfortunately, all known to us previous attempts to develop 
such optimality criterion are either incomplete, or suggest 
the usage of objective functions with the property
that the positions of their minimums change with the shift of the 
point where the beam parameters should be 
reconstructed.\footnote{These suggestions include, for example, the usage of 
the condition number (or some other combination of the singular values) of 
the underlying linear least squares problem as optimality 
criterion.}
It should be clear, that while the usage 
of the objective functions of this sort 
could give useful results in some particular cases,
one hardly can accept any of them as the universal optimality criterion, 
because the results of their optimizations, in general, 
could be completely misleading.

In this and in the accompanying paper~\cite{PartII},
we introduce an optimality criterion which is independent from
the position of the reconstruction point and gives
both, statistical and worst case estimates of the
influence of the beam size measurement errors
on the precision of the reconstruction of the beam parameters.
We use a linear approximation for the beam dynamics 
and assume no coupling between horizontal and vertical motion.

In this paper we develop the geometrical viewpoint on
the dynamics of the second central beam moments, 
which is essential for the understanding 
of the origins of our optimality criterion
and also provides convenient notations for expressing it.
Then we describe standard least squares solution 
of the beam moment reconstruction problem and switch
to the search of invariants connected with the
covariance matrix of the reconstruction errors
(as invariants we mean objects, which are independent from the
position of the reconstruction point).
The optimality criterion itself
and the examples of its application are described 
in the accompanying paper~\cite{PartII}.

\vspace{-0.1cm}
\section{DYNAMICS OF BEAM PARAMETERS FROM
THE GEOMETRICAL VIEWPOINT}

\vspace{-0.1cm}
Let us consider a collection of points in $2$-dimensional 
phase space (a particle beam) and let, for each particle, 
${\mathbf w} =(x, p)^{\top}$
be a vector of canonically conjugate  coordinate $x$ and momentum $p$.
Then, as usual, the beam (covariance) matrix is defined as

\vspace{-0.2cm}
\noindent
\begin{eqnarray}
\Sigma =
(\Sigma_{km}) =
\left\langle
\left({\mathbf w} - \langle {\mathbf w} \rangle \right) 
\cdot 
\left({\mathbf w} - \langle {\mathbf w} \rangle \right)^{\top}
\right\rangle,
\label{Sec2A02}
\end{eqnarray}

\vspace{-0.2cm}
\noindent
where the brackets $\langle \, \cdot \, \rangle$
denote an average over a distribution of the particles in the beam.
Let 

\vspace{-0.2cm}
\noindent
\begin{eqnarray}
A(s_1, s_2) =
\left[
\begin{array}{ll}
a_{11}(s_1, s_2) & a_{12}(s_1, s_2)\\
a_{21}(s_1, s_2) & a_{22}(s_1, s_2)
\end{array}
\right] 
\label{Sec2A0234}
\end{eqnarray}

\vspace{-0.2cm}
\noindent
be a symplectic matrix ($A \in$ Sp(2,R)) which
propagates particle coordinates from the state $s_1$ to the
state $s_2$, i.e let

\vspace{-0.2cm}
\noindent
\begin{eqnarray}
{\mathbf w}(s_2) = A(s_1, s_2) \, {\mathbf w}(s_1).
\label{Sec2A04}
\end{eqnarray}

\vspace{-0.2cm}
\noindent
Then from (\ref{Sec2A02}) and (\ref{Sec2A04}) it follows that
the beam matrix $\Sigma$ evolves between these two
states according to the rule

\vspace{-0.2cm}
\noindent
\begin{eqnarray}
\Sigma(s_2) = A(s_1, s_2) \, \Sigma(s_1) \, A^{\top}(s_1, s_2).
\label{Sec2A05}
\end{eqnarray}

\vspace{-0.2cm}
Let us first extend the domain of the transformation rule (\ref{Sec2A05})
from positive semidefinite symmetric matrices to arbitrary
symmetric matrices and then let us associate with every 
$2 \times 2$ symmetric matrix $\Sigma$ the three component vector

\vspace{-0.2cm}
\noindent
\begin{eqnarray}
{\mathbf m}(\Sigma) = (\Sigma_{11}, \Sigma_{12}, \Sigma_{22})^{\top}.
\label{VecM01}
\end{eqnarray}

\vspace{-0.2cm}
\noindent
With this association the transformation law for the
$2 \times 2$ symmetric matrices (\ref{Sec2A05})
becomes a linear transformation in the 
three dimensional space of ${\mathbf m}$ vectors

\vspace{-0.2cm}
\noindent
\begin{eqnarray}
{\mathbf m}(s_2) = T(s_1, s_2) \, {\mathbf m}(s_1),  
\label{VecM02}
\end{eqnarray}

\vspace{-0.2cm}
\noindent
where the matrix $T = T(A)$ is determined by the relation

\vspace{-0.18cm}
\noindent
\begin{eqnarray}
T(A) =
\left[
\begin{array}{ccc}
a_{11}^2      & 2 a_{11} a_{12}               & a_{12}^2      \\
a_{11} a_{21} & a_{11} a_{22} + a_{12} a_{21} & a_{22} a_{12} \\
a_{21}^2      & 2 a_{21} a_{22}               & a_{22}^2 
\end{array}
\right].
\label{matrxT}
\end{eqnarray}

\vspace{-0.18cm}
\noindent
For an arbitrary $A \in$ Sp(2,R),
the matrix $T(A)$ has unit determinant and all matrices $T$
form a group ($T$-group) of which the symplectic group Sp(2,R)
is the double cover (the matrices $\pm A$ generate the same matrix $T$).
Moreover, an arbitrary matrix $T$ satisfies

\vspace{-0.18cm}
\noindent
\begin{eqnarray}
T^{\top} \,S \,T = S,
\quad
S = 
\left[
\begin{array}{ccc}
0     &  0 & 1 / 2 \\
0     & -1 & 0 \\
1 / 2 &  0 & 0
\end{array}
\right].
\label{matrxC}
\end{eqnarray}

\vspace{-0.18cm}
\noindent
It is a remarkable fact which means that the action of the $T$-group
on ${\mathbf m}$ vectors preserves the symmetric bilinear form

\vspace{-0.2cm}
\noindent
\begin{eqnarray}
B({\mathbf m}_1, \,{\mathbf m}_2) =   
{\mathbf m}_1^{\top} S \,{\mathbf m}_2,
\label{invform01}
\end{eqnarray}

\vspace{-0.2cm}
\noindent
which therefore defines invariant metric.
Because the matrix $S$ has two negative and one positive 
eigenvalues (-1, -1/2, and 1/2), this invariant metric 
is indefinite.\footnote{If, instead of the association law
(\ref{VecM01}), one uses the rule

\vspace{-0.15cm}
\noindent
\begin{eqnarray}
{\mathbf m}(\Sigma) = 
\left((\Sigma_{11} + \Sigma_{22}) \,/\, 2,\,
-\Sigma_{12},\,
(\Sigma_{11} - \Sigma_{22}) \,/\, 2\right)^{\top},
\label{MinkSpace01}
\end{eqnarray}

\vspace{-0.15cm}
\noindent
then one obtains much more known geometry.
The space of ${\mathbf m}$ vectors becomes
the three dimensional Minkowski space
with the standard metric given by the matrix
$S = \mbox{diag}(1,-1,-1)$, and
the $T$-group turns into
the restricted Lorentz group SO$^+$(1,2).
It is clear that both approaches are isomorphic,
but the geometry associated with the rule (\ref{VecM01})
is better suited for our particular purposes.}

The emittance (the invariant norm) of a vector 
$\mathbf{m}=(m_1,m_2,m_3)^{\top}$ 
is defined to be the complex number

\vspace{-0.2cm}
\noindent
\begin{eqnarray}
\varepsilon(\mathbf{m}) = 
\sqrt{\mathbf{m}^{\top} S \,\mathbf{m}}
= 
\sqrt{m_1 m_3 - m_2^2},
\label{invform02}
\end{eqnarray}

\vspace{-0.2cm}
\noindent
where $\varepsilon(\mathbf{m})$ is either positive, zero,
or positive imaginary.

In the following we will say that the vector ${\mathbf m}$ is beamlike, 
if the corresponding to it symmetric $2 \times 2$ matrix $\Sigma$
is positive definite, i.e. if the first component $m_1$ of the vector
${\mathbf m}$ and its emittance $\varepsilon(\mathbf{m})$ are both positive.
Note that if $\mathbf{m}_1$ and $\mathbf{m}_2$ are two beamlike vectors,
then

\vspace{-0.2cm}
\noindent
\begin{eqnarray}
\mathbf{m}_1^{\top} S \,\mathbf{m}_2
\geq 
\varepsilon(\mathbf{m}_1) \, \varepsilon(\mathbf{m}_2),
\label{invform03}
\end{eqnarray}

\vspace{-0.2cm}
\noindent
which is the reverse Cauchy-Bunyakovsky-Schwarz inequality.
Moreover, the two sides in (\ref{invform03}) are equal if and only if 
$\mathbf{m}_1$ and $\mathbf{m}_2$ are two proportional vectors.

So we have obtained the following geometric picture.
The $2 \times 2$ symmetric matrices are put into one to one correspondence
with the points of the three dimensional indefinite metric space,
where the nondegenerated beam matrices occupy the convex
region for which the nonnegative ($m_1 \geq 0$)
part of the conical surface $\varepsilon^2(\mathbf{m}) = 0$
is the boundary. Under the action of the $T$-group this
convex region splits into a set of the positive ($m_1 > 0$) sheets  
of the two-sheeted hyperboloids  
$\varepsilon^2(\mathbf{m}) = const > 0$  (orbits),
and on each orbit the $T$-group acts 
transitively (see Fig.1).

\begin{figure}[ht]
    \centering
    \includegraphics*[width=25mm]{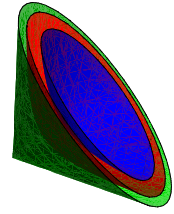}
    \vspace{-0.4cm}
    \caption{Boundary of the set of the beamlike vectors (green)
    and two invariant orbits (red and blue) inside it.}
    \label{fig1}
    \vspace{-0.5cm}
\end{figure}

If the emittance of a beamlike vector $\mathbf{m}$ is known, 
then the dynamics of this vector is completely determined by the behavior
of its projection onto the special 
orbit ${\cal T}_s$ labeled by the emittance equal to one (Twiss surface), i.e.
by the dynamics of the Twiss vector and Twiss parameters

\vspace{-0.2cm}
\noindent
\begin{eqnarray}
\mathbf{t}(\mathbf{m}) = 
(\beta(\mathbf{m}), -\alpha(\mathbf{m}), \gamma(\mathbf{m}))^{\top}
\stackrel{\rm def}{=} \mathbf{m} \,/\, \varepsilon(\mathbf{m}).
\label{mp01}
\end{eqnarray}

\vspace{-0.2cm}
\noindent
And as the next step in the development of the geometrical view
on the dynamics of the beam parameters, 
let us consider the Twiss surface with the metric
induced from the ambient metric (\ref{invform01}).
It is possible to show that it is a model of the 
hyperbolic Lobachevsky plane. A positive outcome from this fact is that
the distance between the Twiss vectors can be measured 
using the hyperbolic distance function 

\vspace{-0.2cm}
\noindent
\begin{eqnarray}
d_H(\mathbf{t}_1, \mathbf{t}_2)
=
\mbox{arccosh}(m_p(\mathbf{t}_1, \mathbf{t}_2)),
\label{mp03}
\end{eqnarray}

\vspace{-0.2cm}
\noindent
where 

\vspace{-0.5cm}
\noindent
\begin{eqnarray}
m_p(\mathbf{t}_1, \mathbf{t}_2)
= \mathbf{t}_1^{\top} S \, \mathbf{t}_2
\label{mp02}
\end{eqnarray}

\vspace{-0.2cm}
\noindent
is the betatron mismatch parameter.
Note that if the difference $\,m_p - 1\,$ is small, then

\vspace{-0.2cm}
\noindent
\begin{eqnarray}
d_H =
\sqrt{2 (m_p - 1)}
\cdot
\big(\,
1 - (m_p - 1)\, /\, 12 + \ldots
\,\big).
\label{mp033}
\end{eqnarray}

\vspace{-0.2cm}
Let us give here a brief summary of the most interesting 
outcomes of this section.
First, it is the important role of the invariant bilinear
form (\ref{invform01}), which is the origin of both,
beam emittance and betatron mismatch parameter. 
So, it should be no surprise, when the matrix of this form
will regularly show itself during the course of this paper
and will also enter our final optimality criterion.
Then, we have seen that there is a function of the
betatron mismatch parameter which is better suited for
the comparison of the Twiss vectors, than the mismatch
parameter itself. It is the
hyperbolic distance function (\ref{mp03}).
Besides that, we hope that the geometrical interpretation
of the dynamics of the beam matrices has shown
more clearly that, in order to compare two beamlike vectors
in invariant manner,
we have to look at two different quantities, at the difference
of their emittances and at the hyperbolic distance
(or mismatch) between their Twiss parameters. 
It doesn't seem that there exists any 
``natural way" to unite these two quantities into a single
value, which, in the next turn, means that
the optimality criterion, which we are looking for,
should be a vector criterion and
should contain two different objective functions.

\vspace{-0.1cm}
\section{USAGE OF LEAST SQUARES FOR BEAM MOMENT
RECONSTRUCTION}

\vspace{-0.1cm}
Let us assume that the beam size was measured
in the $n$ states $s_1, \ldots, s_n$
and let $T(r, s_m)$ be a matrix which transport the 
$\mathbf{m}$ vectors
from the reconstruction state $r$ to the $m$-th measurement 
state $s_m$.
If $\mathbf{m}_0(r)= \varepsilon_0 \,\mathbf{t}_0(r)$ 
is the beamlike vector matched to the
measurements system, then 

\vspace{-0.2cm}
\noindent
\begin{eqnarray}
\mathbf{b}_0 = M(r) \,\mathbf{m}_0(r),
\label{matrxM098}
\end{eqnarray}

\vspace{-0.2cm}
\noindent
where

\vspace{-0.2cm}
\noindent
\begin{eqnarray}
M(r) =
\left[
\begin{array}{ccc}
T_{11}(r, s_1) & T_{12}(r, s_1) & T_{13}(r, s_1) \\
\vdots         & \vdots         & \vdots         \\
T_{11}(r, s_n) & T_{12}(r, s_n) & T_{13}(r, s_n) 
\end{array}
\right],
\label{matrxM}
\end{eqnarray}

\vspace{-0.2cm}
\noindent
is the vector of the squares of the rms beam widths
as they actually are in the states $s_1, \ldots, s_n$.

Unfortunately, the measurement system does not
deliver us the vector $\mathbf{b}_0$, but gives us
instead the vector

\vspace{-0.2cm}
\noindent
\begin{eqnarray}
\mathbf{b}_{\varsigma} = \mathbf{b}_0 + \varsigma,
\label{SEC1_3}
\end{eqnarray}

\vspace{-0.2cm}
\noindent
where 
$\varsigma = (\varsigma_1, \ldots, \varsigma_n)^{\top}$ 
is the vector of the measurement errors.
In the following we will assume that the vector $\varsigma$
is random from measurement to measurement and
(over many measurements)
has zero mean and positive definite covariance matrix,
i.e. that

\vspace{-0.2cm}
\noindent
\begin{eqnarray}
\big< \varsigma \big> =  0, 
\;\;\;\;\;
V_{\varsigma} = 
\big< \varsigma \varsigma^{\top} \big>
- 
\big< \varsigma \big> 
\big< \varsigma \big>^{\top}
> 0, 
\label{SEC1_4}
\end{eqnarray}

\vspace{-0.2cm}
\noindent
where now and later on the brackets $\langle \, \cdot \, \rangle$
mean an average over the measurement 
statistics.\footnote{The matrix $V_{\varsigma}$ can
be a function of the vector $\mathbf{b}_0$, 
i.e. the measurement errors can depend on the measured beam sizes.}

Let us assume that the numerical value of the matrix $V_{\varsigma}$
is known, and let us take as an estimate 
$\mathbf{m}_{\varsigma}(r)$
of the vector $\mathbf{m}_0(r)$ in the presence of the measurement errors
solution of the following weighted linear least squares 
problem\footnote{Note that the weight matrix in (\ref{SEC1_9}) can be taken
different from $V_{\varsigma}^{-1}$. 
It will complicate the formula (\ref{SEC1_16}), but most of our general
results will stay unaltered.}

\vspace{-0.2cm}
\noindent
\begin{eqnarray}
\min_{\mathbf{m}_{\varsigma}(r)}
\left( 
M \mathbf{m}_{\varsigma} - \mathbf{b}_{\varsigma} 
\right)^{\top}
V_{\varsigma}^{-1}
\left( 
M \mathbf{m}_{\varsigma} - \mathbf{b}_{\varsigma} 
\right).
\label{SEC1_9}
\end{eqnarray}

\vspace{-0.2cm}
\noindent
The problem (\ref{SEC1_9}) always has solutions 
and, if we will assume that the matrix $M$ has
full column rank, 
then the solution is unique 
and is given by the formula

\vspace{-0.2cm}
\noindent
\begin{eqnarray}
\mathbf{m}_{\varsigma}(r) =
\left[M^{\top}(r) V_{\varsigma}^{-1} M(r) \right]^{-1}
M^{\top}(r)\, V_{\varsigma}^{-1} \,\mathbf{b}_{\varsigma}.
\label{SEC1_15}
\end{eqnarray}

\vspace{-0.2cm}
Note that the important condition for the matrix $M$
to have full column rank is equivalent to the property
of the determinant of the matrix $M^{\top} V_{\varsigma}^{-1} M$
to be nonzero. If we assume that the matrix $V_{\varsigma}$
is a diagonal matrix 

\vspace{-0.2cm}
\noindent
\begin{eqnarray}
V_{\varsigma} =
\mbox{diag}
\left(
\sigma_1^2 ,\, \sigma_2^2 ,\, \ldots ,\, \sigma_n^2\,
\right)
\label{SEC1_17}
\end{eqnarray}

\vspace{-0.2cm}
\noindent
with all $\sigma_m > 0$, 
then the expression for this determinant can be obtained 
in the explicit form as follows

\vspace{-0.2cm}
\noindent
\begin{eqnarray}
\Delta_{\varsigma} =
\det \left[ M^{\top}(r) V_{\varsigma}^{-1} M(r) \right] 
\nonumber
\end{eqnarray}

\vspace{-0.5cm}
\noindent
\begin{eqnarray}
=
\frac{2}{3} \sum_{i,j,k = 1}^{n} 
\frac{a_{12}^2(s_i, s_j)}{\sigma_i\,\sigma_j}
\cdot
\frac{a_{12}^2(s_j, s_k)}{\sigma_j\,\sigma_k}
\cdot
\frac{a_{12}^2(s_k, s_i)}{\sigma_k\,\sigma_i}.
\label{matrxM3}
\end{eqnarray}

\vspace{-0.3cm}
\section{INVARIANTS CONNECTED WITH\\
THE COVARIANCE MATRIX\\
OF RECONSTRUCTION ERRORS}

\vspace{-0.1cm}
The calculation of the covariance matrix of the errors 
of the estimate (\ref{SEC1_15}) is standard
and gives the following result

\vspace{-0.15cm}
\noindent
\begin{eqnarray}
V_m(r) = 
\langle \tilde{\mathbf{m}}_{\varsigma}(r)\, 
\tilde{\mathbf{m}}_{\varsigma}^{\top}(r) \rangle
= 
\left[ M^{\top}(r) V_{\varsigma}^{-1} M(r) \right]^{-1},
\label{SEC1_16}
\end{eqnarray}

\vspace{-0.15cm}
\noindent
where $\tilde{\mathbf{m}}_{\varsigma}$ is the error vector
given by the equality

\vspace{-0.2cm}
\noindent
\begin{eqnarray}
\tilde{\mathbf{m}}_{\varsigma}(r)
=
\mathbf{m}_{\varsigma}(r) -
\mathbf{m}_0(r).
\label{SEC1_31}
\end{eqnarray}

\vspace{-0.2cm}
Let $T(r_1, r_2)$ be a matrix which transport 
$\mathbf{m}$ vectors from the state $s = r_1$  
to the state $s = r_2$. Because

\vspace{-0.2cm}
\noindent
\begin{eqnarray}
M(r_2) = M(r_1)\, T^{-1}(r_1, r_2),
\label{matr3}
\end{eqnarray}

\vspace{-0.2cm}
\noindent
one can show that,
when the position of the reconstruction point changes,
the vector $\mathbf{m}_{\varsigma}$
propagates as any other $\mathbf{m}$ vector

\vspace{-0.4cm}
\noindent
\begin{eqnarray}
\mathbf{m}_{\varsigma}(r_2)
= T(r_1, r_2)\,
\mathbf{m}_{\varsigma}(r_1),
\label{SEC1_31DDD}
\end{eqnarray}

\vspace{-0.2cm}
\noindent
and the matrix $V_m$ evolves according to the congruence 

\vspace{-0.18cm}
\noindent
\begin{eqnarray}
V_m(r_2) = T(r_1, r_2)\, V_m(r_1) \,T^{\top}(r_1, r_2).
\label{matr37}
\end{eqnarray}

\vspace{-0.18cm}
\noindent
Multiplying both sides of the equation (\ref{matr37}) from 
the right hand side by the matrix $S$ and using 
the identity (\ref{matrxC}),  we turn the congruence
(\ref{matr37}) into the similarity transformation

\vspace{-0.18cm}
\noindent
\begin{eqnarray}
\left[V_m(r_2) S\right] = T(r_1, r_2)\, \left[V_m(r_1) S \right]\, T^{-1}(r_1, r_2),
\label{matr39}
\end{eqnarray}

\vspace{-0.18cm}
\noindent
which means that the eigenvalues of the matrix $V_m S$
are invariants, i.e. they are independent from the
position of the reconstruction point.
Because

\vspace{-0.2cm}
\noindent
\begin{eqnarray}
V_m S =
V_m^{1/2} 
\left(
V_m^{1/2} 
S\,
V_m^{1/2} 
\right)
V_m^{-1/2}, 
\label{ev1009}
\end{eqnarray}

\vspace{-0.2cm}
\noindent
these eigenvalues are real numbers
and the inertias of the matrices $V_m S$ and $S$ coincide,
i.e the matrix $V_m S$ has one positive and two negative
eigenvalues which in the following we will denote as 

\vspace{-0.2cm}
\noindent
\begin{eqnarray}
\lambda_1 > 0 > \lambda_2 \geq \lambda_3.
\label{ev10097}
\end{eqnarray}

\vspace{-0.2cm}
\noindent
If the errors in the beam size determination
at different measurement states
can be considered as uncorrelated 
(i.e. if the matrix $V_{\varsigma}$ is
diagonal), then, in addition to the inequalities
(\ref{ev10097}), the following properties hold:

\vspace{-0.2cm}
\noindent
\begin{eqnarray}
1 \,/\,\lambda_1 + 1\,/\,\lambda_2 + 1\,/\,\lambda_3 = 0 
\label{vif0198}
\end{eqnarray}

\vspace{-0.2cm}
\noindent
and

\vspace{-0.2cm}
\noindent
\begin{eqnarray}
\lambda_1 + \lambda_2 + \lambda_3 =
-\frac{1}{2 \Delta_{\varsigma}} \sum_{i,j = 1}^{n} 
\left(\frac{a_{12}^2(s_i, s_j)}
{\sigma_i \,\sigma_j}
\right)^2 < 0.
\label{matrxM3786}
\end{eqnarray}

\vspace{-0.2cm}
The eigenvalues of the matrix $V_m S$ do not
exhaust all invariants connected with the
covariance matrix $V_m$. 
Using the formulas (\ref{matr3}) and (\ref{matr37}),
and the transformation rule for the Twiss vectors
(which is the same as for any other $\mathbf{m}$ vectors),
one can show that the quadratic forms

\vspace{-0.2cm}
\noindent
\begin{eqnarray}
{\cal F} = 
\mathbf{t}_0^{\top} S\, V_m S \,\mathbf{t}_0,
\label{INVQF01}
\end{eqnarray}

\vspace{-0.2cm}
\noindent
\begin{eqnarray}
{\cal G} = 
\mathbf{t}_0^{\top}  V_m^{-1} \,\mathbf{t}_0,
\label{INVQF02}
\end{eqnarray}

\vspace{-0.2cm}
\noindent
and the matrices

\vspace{-0.2cm}
\noindent
\begin{eqnarray}
U = M V_m S\, V_m M^{\top},
\label{INVQF03}
\end{eqnarray}

\vspace{-0.2cm}
\noindent
\begin{eqnarray}
W = M V_m S\, \mathbf{t}_0\, \mathbf{t}_0^{\top} S \,V_m M^{\top}
\label{INVQF04}
\end{eqnarray}

\vspace{-0.2cm}
\noindent
are invariants, i.e. the values of the 
quadratic forms ${\cal F}$ and ${\cal G}$,
as well as the elements of the matrices $U$ and $W$
are all independent from the choice of the
position of the reconstruction point.

\end{document}